\documentclass[onecolumn,aps,prl,superscriptaddress,showpacs,secnumroman,showkeys]{revtex4}

\usepackage{amsmath,bm,CJK}%
\usepackage{graphicx}%
\usepackage{color}%
\usepackage{float}
\usepackage{booktabs}
\usepackage{supertabular}
\begin{document}

\preprint{}

\title{Measurement of the $^{159}$Tb(n, $\gamma$) cross section at the CSNS Back-n facility}


\author{S. Zhang}
\email[E-mail at:]{zsylt@imun.edu.cn}
\affiliation{College of Mathematics and Physics, Inner Mongolia Minzu University, Tongliao 028000, China.}
\author{G. Li}
\affiliation{College of Mathematics and Physics, Inner Mongolia Minzu University, Tongliao 028000, China.}
\author{W. Jiang}
\affiliation{Institute of High Energy Physics, Chinese Academy of Sciences, Beijing 100049, China.}
\author{D.X. Wang}
\affiliation{College of Mathematics and Physics, Inner Mongolia Minzu University, Tongliao 028000, China.}
\author{J. Ren}
\affiliation{China Institute of Atomic Energy, Beijing 102413, China.}
\author{E.T. Li}
\affiliation{Institute for Advanced Study in Nuclear Energy $\&$ Safety, College of Physics and Optoelectronic Engineering, Shenzhen University, Shenzhen 518060, China.}
\author{M. Huang}
\email[E-mail at:]{huangmeirong@imun.edu.cn}
\affiliation{College of Mathematics and Physics, Inner Mongolia Minzu University, Tongliao 028000, China.}
\author{J.Y. Tang}
\affiliation{Institute of High Energy Physics, Chinese Academy of Sciences, Beijing 100049, China.}
\author{X.C. Ruan}
\affiliation{China Institute of Atomic Energy, Beijing 102413, China.}
\author{H.W. Wang}
\affiliation{Shanghai Advanced Research Institute, Chinese Academy of Sciences, Shanghai 201210, China.}
\author{Z.H. Li}
\affiliation{China Institute of Atomic Energy, Beijing 102413, China.}
\author{Y.S. Chen}
\affiliation{China Institute of Atomic Energy, Beijing 102413, China.}
\author{L.X. Liu}
\affiliation{Shanghai Advanced Research Institute, Chinese Academy of Sciences, Shanghai 201210, China.}
\author{X.X. Li}
\affiliation{Shanghai Institute Applied Physics, Chinese Academy of Sciences, Shanghai 201800, China.}
\author{Q.W. Fan}
\affiliation{China Institute of Atomic Energy, Beijing 102413, China.}
\author{R.R. Fan}
\affiliation{Institute of High Energy Physics, Chinese Academy of Sciences, Beijing 100049, China.}
\author{X.R. Hu}
\affiliation{Shanghai Institute Applied Physics, Chinese Academy of Sciences, Shanghai 201800, China.}
\author{J.C. Wang}
\affiliation{College of Mathematics and Physics, Inner Mongolia Minzu University, Tongliao 028000, China.}
\affiliation{China Institute of Atomic Energy, Beijing 102413, China.}
\author{X. Li}
\affiliation{College of Mathematics and Physics, Inner Mongolia Minzu University, Tongliao 028000, China.}
\author{D.D. Niu}
\affiliation{College of Mathematics and Physics, Inner Mongolia Minzu University, Tongliao 028000, China.}
\author{N. Song}
\affiliation{College of Mathematics and Physics, Inner Mongolia Minzu University, Tongliao 028000, China.}
\affiliation{China Institute of Atomic Energy, Beijing 102413, China.}
\author{M. Gu}
\affiliation{College of Mathematics and Physics, Inner Mongolia Minzu University, Tongliao 028000, China.}



\begin{abstract}
The stellar (n, $\gamma$) cross section data for the mass numbers around A $\approx$ 160 are of key importance to nucleosynthesis
in the main component of the slow neutron capture process, which occur in the thermally pulsing asymptotic giant branch (TP--AGB).
The new measurement of (n, $\gamma$) cross sections for $^{159}$Tb was performed using the C$_6$D$_6$ detector system at the back streaming
white neutron beam line (Back-n) of the China spallation neutron source (CSNS) with neutron energies ranging from 1 eV to 1 MeV.
Experimental resonance capture kernels were reported up to 1.2 keV neutron energy with this capture measurement. Maxwellian-averaged cross sections (MACS) were derived from the measured $^{159}$Tb (n, $\gamma$) cross sections at $kT$ = 5 $\sim$ 100 keV and
are in good agreement with the recommended data of KADoNiS-v0.3 and JEFF-3.3,
while KADoNiS-v1.0 and ENDF-VIII.0 significantly overestimate the present MACS up to 40$\%$ and 20$\%$, respectively. A sensitive test of the s-process nucleosynthesis was also
performed with the stellar evolution code MESA. Significant changes in abundances around A $\approx$ 160 were observed between the ENDF/B-VIII.0 and present measured rate of $^{159}$Tb(n, $\gamma$)$^{160}$Tb in the MESA simulation.
\end{abstract}


\maketitle


\section{I. INTRODUCTION}
The abundances of elements heavier than iron in the solar system are mainly produced by the slow neutron capture process (s-process)~\cite{Reifarth2014} and the rapid neutron capture process (r-process)~\cite{Thielemann2011} in stars as found by Burbidge et al.~\cite{Burbidge1957} and Cameron~\cite{Cameron1957} as early as in 1957. Almost less than 1\% of heavy elements are ascribed to the production of photo-disintegration processes (the so-called p/$\gamma$ process), neutrinos, and charged-particle induced reactions~\cite{Arnould2003,Woosley1990}. The s-process takes place during stellar evolution and path through nuclei along the valley of $\beta$ stability with lower neutron densities and temperatures. The weak s-process is responsible for producing isotopes up to A $\approx$ 90 and occurs during He and C burning in massive stars. While the main s-process contributes isotopes A $\approx$ 90-208 and is produced in the thermally pulsing asymptotic giant branch (TP-AGB) phase of low- and intermediate-mass stars. The r-process is related to explosive nucleosynthesis in massive stars and binary star mergers with relatively high neutron densities.

\begin{figure}[h]
\centering
\includegraphics[scale=0.55]{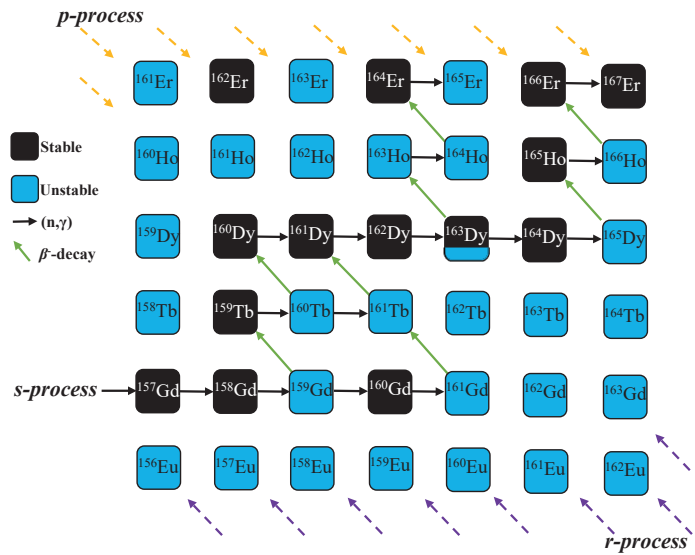}
\caption{\footnotesize (color online) The s-process reaction path in the region of terbium. Black solid boxes indicate stable; blue empty boxes indicate unstable isotopes. Arrows to the right and the next higher elements represent neutron capture reactions and $\beta^{-}$ decays, respectively. The isotope$^{163}$Dy is a terrestrially stable nucleus, which becomes unstable at stellar temperature~\cite{FVoss1999prc}.}
\label{Fig1}
\end{figure}

Terbium is thought to be mainly produced by the explosive r-process~\cite{Thielemann2011}, while about 9\% is made by main s-process in low- and intermediate-mass TP-AGB stars~\cite{Bisterzo2011}. The s-process reaction path around terbium is sketched in Fig. \ref{Fig1}. The $^{160}$Tb is an important branching point, which is shielded against the $\beta^{-}$ decay chains from the r-process by stable nuclei $^{160}$Gd. It will affect the s-only isotope $^{160}$Dy considered as an reference point for the reaction flow. Also, it will contribute to the production of p-nuclei $^{164}$Er through the temperature sensitive branching at $^{163}$Dy to $^{163}$Ho by the nuclear reaction path $^{163}$Ho(n,$\gamma$)$^{164}$Ho($\beta$)$^{164}$Er in the stellar environment~\cite{FVoss1999prc,ProcessSensitiv2006apj,Hayakawa2008apj}. 


Two groups of experimental neutron capture cross section data for $^{159}$Tb(n, $\gamma$) reaction  were reported between 1961 and 1991~\cite{Block1961,Gibbons1961,Lpine1972,Mizumoto1978,Ohkubo1979,Bokhovko1991} in the astrophysical interested energy region. Data from Block et al.~\cite{Block1961}, Gibbons et al.~\cite{Gibbons1961}, L\'{e}pine et al.~\cite{Lpine1972} are significantly larger than data of Mizumoto et al.~\cite{Mizumoto1978}, Ohkubo et al.~\cite{Ohkubo1979} and Bokhovko et al.~\cite{Bokhovko1991}. 
The recommended Maxwellian-averaged cross sections (MACS) of $^{159}$Tb(n, $\gamma$) at stellar temperature $kT$ = 30 keV is 1580 $\pm$ 150 mb in the KADoNiS-v0.3~\cite{Dillmann} database, which is widely used in stellar nucleosynthesis calculations, derived from the neutron capture cross section data of Mizumoto et al.~\cite{Mizumoto1978} and Bokhovko et al.~\cite{Bokhovko1991}. While, this value reported by the recent activation measurement~\cite{Praena2014} is 2166 $\pm$ 181 and varies from 1116 to 2949 mb in the collection list of KADoNiS-v1.0~\cite{kadonisV10} and references therein. Because of so large discrepancies, more new measurements of stellar neutron capture cross section of  $^{159}$Tb are desired for improving the reliability  of recommended data in KADoNiS-v1.0~\cite{kadonisV10}. As well, much higher accurate neutron capture cross sections for the terbium isotopes are required to s-process nucleosynthesis, and separation of the r and s components in observed Tb abundance through subtraction of the s abundances from the respective solar values. In addition, the $^{159}$Tb(n, $\gamma$) cross section data is very important for the design of nuclear reactor and the research in nuclear structure~\cite{Chen1988,Dzysiuk2015npa}.

In the past years, a series of neutron radiative capture experiments have been performed using C$_6$D$_6$ detectors with the total energy technique~\cite{Tm2020prl,Lederer2019,181Ta2017,Ni2013prl,Sm2004prl} and total absorption calorimeter consists of BaF$_{2}$ scintllator arrays~\cite{65Cu2019,Roig2016,Wisshak2006,Wisshak1990} in combination with the $^{7}$Li(p, n)$^{7}$Be reaction neutrons and white neutron sources. 
Both C$_6$D$_6$ and BaF$_{2}$ detection systems, respectively, were constructed at the back streaming white neutron facility (Back-n) of the China Spallation Neutron Source(CSNS)~\cite{Tang2021,Tang20212}, which is the only high intense pulsed spallation reaction neutron source in China at present. The main physics motivation of these systems is measuring the neutron capture data related to the nuclear astrophysics, advanced nuclear energy technologies such as accelerator-driven subcritical systems, thorium-based molten salt reactor, and fourth generation reactors. The detection systems, experimental techniques, background study and resonance energy region measurements have been published in Refs~\cite{Ren2022cpc,Ren2021nima,Li2021,2021GTAF}.
In this work, we report the new experiment of neutron capture cross sections for $^{159}$Tb over the energy range from 1 eV to 1 MeV via the time-of-flight (TOF) method combined with the white neutron beam at the CSNS Back-n facility. The resonance kernels of $^{159}$Tb (n, $\gamma$) were obtained up to 1.2 keV, and the MACS for the astrophysical interest energy region from $kT$ = 5 $\sim$ 100 keV were calculated.

\section{II. Experiment}
The measurement of the $^{159}$Tb (n, $\gamma$) was performed at the CSNS Back-n facility~\cite{Tang2021,Tang20212} in China. Neutrons were produced via the spallation reactions by bombarding a massive tungsten target with a 1.6 GeV proton beam, operating at a 25 Hz repetition rate, and collimated to a 30 mm diameter beam spot at the sample position. The neutron energies ranging from thermal to a few hundred MeV were determined using the time of flight (TOF) method with a 76 m long flight path. The neutron capture yields were studied using the total energy principle based on an array of four C$_6$D$_6$ detectors with the pulse height weighting techniques (PHWT)~\cite{Tain2002,Abbond2004,Schill2012}.
The neutron flux in the energy range of 1 eV to 100 MeV was obtained with a combination of the data from dedicated measurements of $^{6}$LiF-silicon (Li-Si) detector array~\cite{liqiang2019} and $^{235}$U loaded multilayer fission chamber~\cite{chenyh2019}. The Li-Si detector neutron flux was normalized to $^{235}$U data at 10 $\sim$ 20 keV since its position is about 20 m away from the sample.
The samples used for this experiment are listed in Table \ref{tab1}. A metallic terbium sample of natural abundance with a thickness of 0.2 mm was used for the $^{159}$Tb (n, $\gamma$) reaction study. The $^{197}$Au sample was used as the standard for calculating the relative cross section of $^{159}$Tb (n, $\gamma$) in neutron energies ranging from 2 keV to 1 MeV.
The $^{nat}$Pb samples were used to normalize the neutron fluence and to evaluate in-beam $\gamma$-rays and scattered-neutron backgrounds, respectively. The empty sample was also run to assess the background produced from the upstream devices, such as the sample holders, etc.
All experimental data were collected on an event-by-event basis high-performance data acquisition system with a 12-bit full-waveform
digitizers sampling at 1 GS/s~\cite{Wang2018}. The offline data analysis was done on the CERN ROOT framework~\cite{ROOT}.

\begin{table}[htbp]\small
\centering
\caption{\small Characteristic parameters of samples}
\label{tab1}
\renewcommand\arraystretch{0.95}
\setlength{\tabcolsep}{1.5mm}{
\begin{tabular}{ccccccccc}
\toprule
\hline
Sample & Thickness (mm)  &  Diameter (mm)  &  Mass (mg)  &  Area desity ($atom\cdot b^{-1}$)\\
\midrule
\hline
$^{nat}$Tb & 0.20    &  30     & 1169.14& $6.27\times 10^{-4}$\\
$^{197}$Au & 0.10     &  30     & 1357.17& $5.87\times 10^{-4}$\\
$^{nat}$Pb & 0.53    &  30     & 4249.75& $1.75\times 10^{-3}$\\
Empty holder &  &  &  &  &\\
\bottomrule
\hline
\end{tabular}
}
\end{table}

\section{III. Data analysis and results}
The incident neutron energy $E_n$ was determined by employing the TOF method through the nonrelativistic kinematics formula (\ref{eq1}). The neutron effective flight path $L$ was obtained by analyzing the low energy resonances of the gold sample.
\begin{normalsize}
\begin{equation}
E_n = (\frac{72.2977\times L}{t_n})^2
\label{eq1}
\end{equation}
\end{normalsize}
where $E_n$ is in MeV, $L$ is in meters, and $t_n$ is the flight time in ns.

The experimental neutron capture yield as a function of $E_n$ can be calculated as:
\begin{normalsize}
\begin{equation}
Y_{exp}(E_n)=\frac{1}{f}\frac{C^w(E_n)-B^w(E_n )}{\Phi(E_n)\times E_c}
\label{eq2}
\end{equation}
\end{normalsize}
where $f$ is the normalization factor determined by self-normalizing the measured capture yield of 4.9 eV resonance of $^{197}$Au and 11.1 eV resonance of $^{159}$Tb, respectively, based on the saturated resonance technique~\cite{NIM1979,NIMA2007}. $E_c$ is the detection efficiency for a capture event, $\Phi(E_n$) is the neutron flux spectrum, and $C^w(E_n)$ and $B^w(E_n)$ are the weighted count spectrum and the total background spectrum, respectively.

The PHWT method is essential for the $(n, \gamma$) cross section measurement using the C${_6}$D${_6}$ detection system. Its function is to make the detection efficiency $\varepsilon_{\gamma}$ proportional to the incident $\gamma$ ray energy $E_\gamma$, as Eq. (\ref{eq3})
\begin{normalsize}
\begin{equation}
\varepsilon_{\gamma}=\sum\nolimits_{i=1}^{n}WF_i(E_{d}) R(E_{d},E_\gamma)=\alpha E_\gamma
\label{eq3}
\end{equation}
\end{normalsize}
where $\alpha$ is the constant parameter, $R$ is the detector response function, $E_{d}$ is the energy bin of pulse height spectrum, and WF is the weighting function, which can be approximated by a 5$^{th}$ polynomial function. The polynomial coefficients are obtained with a minimum least-squares fit to the the detector response for 27 different monoenergetic $\gamma$ rays from 0.1 MeV to 10 MeV simulated by the GEANT4 code~\cite{Agostinelli2003}, taking into account the detailed model of the detection systems and the sample.

The background consists of two components, including sample-independent and sample-dependent. The former was evaluated by sample-out runs under the same experimental conditions. The sample-dependent background was mainly induced by in-beam $\gamma$ rays and scattered neutrons on the sample. Its contributions were determined by measurements with a $^{nat}$Pb sample and parameterized with eq.(\ref{eq4}).
\begin{normalsize}
\begin{equation}
B(E_n) = f_\gamma B_\gamma (E_n) + f_n B_n(E_n)
\label{eq4}
\end{equation}
\end{normalsize}
where the $B_\gamma$ and $B_n$ denote the background contribution of in-beam gamma rays and scattered neutrons, respectively, and can be formulated as eqs. (\ref{eq5}-\ref{eq6}).
\begin{normalsize}
\begin{equation}
B_\gamma(E_n) = b\times e^{\frac{-c}{\sqrt{E_n}}} + d\times e^{-e\times \sqrt{E_n}} + f
\label{eq5}
\end{equation}
\end{normalsize}

\vspace{-0.5cm}
\begin{normalsize}
\begin{equation}
B_n(E_n) = \frac{a}{\sqrt{E_n}}
\label{eq6}
\end{equation}
\end{normalsize}

\begin{figure}[h!]
\centering
\includegraphics[scale=0.6]{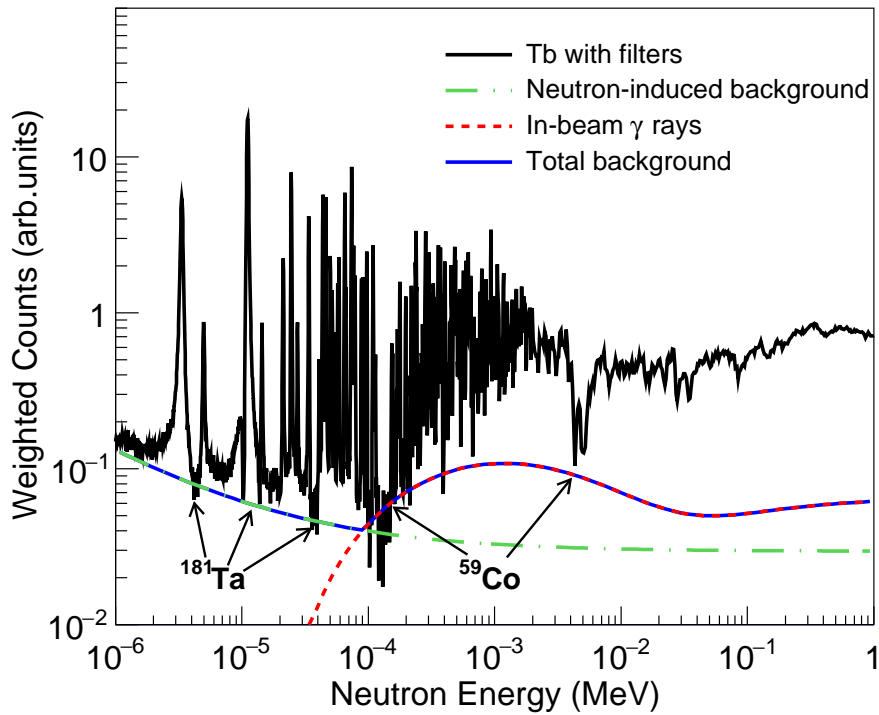}
\caption{\footnotesize (color online) $^{159}$Tb(n, $\gamma$) capture yield with filters and background components are shown as a function of neutron energy. Black and blue solid lines indicate the $^{159}$Tb sample with $^{181}$Ta + $^{59}$Co neutron filters and total background which is made of scattered neutrons (green dash-dotted lines) and in-beam $\gamma$ rays (red dashed lines).}
\label{Fig2}
\end{figure}

In order to normalize these background components, $B_\gamma$ and $B_n$, dedicated runs with $^{181}$Ta and $^{59}$Co neutron filters were performed. Fig. \ref{Fig2} shows the energy spectrum obtained from terbium with filters and background components. The normalization factors $f{_n}$ and $f{_\gamma}$ for $B_n$ and $B_\gamma$ components were obtained by matching the filter dips.
The effect of the filters on in-beam gamma rays and neutrons was analyzed by considering the neutron flux and energy distribution of in-beam gamma rays. The neutron and gamma energy spectra of Back-n were sampled as incident particle energy spectra using GEANT4 code to simulate with and without filters, and the counts of scattered neutrons and gamma were recorded at the detector position~\cite{Ren2022cpc}. The neutron and gamma attenuation factors are 0.92 and 0.68, respectively, and are used for corrections to $f{_n}$ and $f{_\gamma}$.

\begin{figure}[h!]
\vspace{-10pt}
\centering
\includegraphics[scale=0.47]{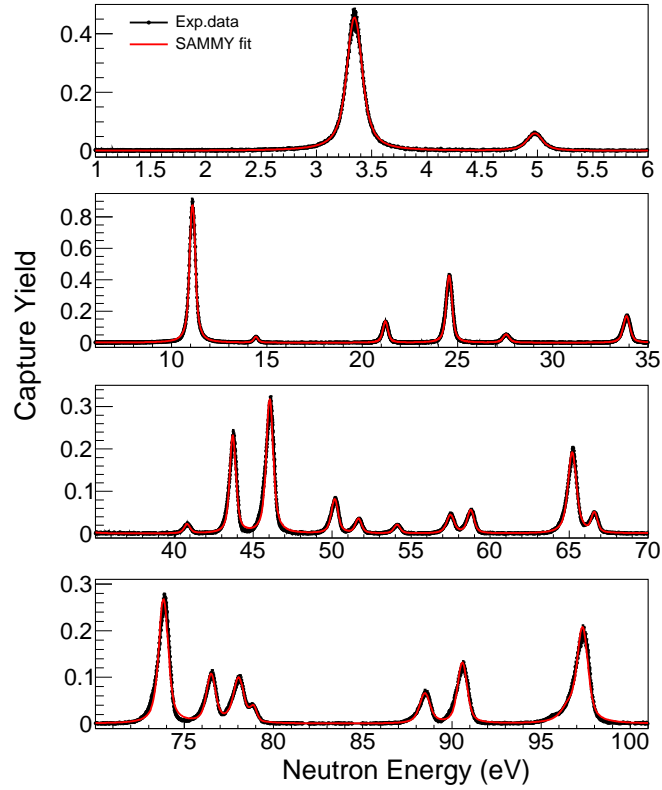}
\vspace{-0.5cm}
\caption{\footnotesize (color online) The SAMMY fits to the experimental capture yields of $^{159}$Tb (n, $\gamma$).}
\label{Fig3}
\end{figure}

The resonance parameters (i.e., the radiative width $\Gamma_\gamma$ and the neutron width $\Gamma_n$) were extracted using the R-matrix code SAMMY~\cite{Larson2008} fit to the experimental capture yields of $^{159}$Tb(n, $\gamma$) in the resonance region up to 1.2 keV using the JEFF-3.3~\cite{JEFF} evaluated data as initial values. The code includes the experimental conditions, such as sample composition, multiple scattering, neutron self-shielding, and the Doppler effect at room temperature. Also, the energy-dependent parameterized resolution function of the Back-n facility~\cite{BingJiang} was used to broaden the resonance peaks. The resonance parameters obtained in this work were compared with those from JEFF-3.3 evaluations in Appendix A.
The experimental kernels $k$ $( k = g \Gamma_n \Gamma_\gamma/(\Gamma_n + \Gamma_\gamma)$, $g$ is the statistical factor) highly agreed with the evaluations below 100 eV, while some differences were observed in the energy range from 100 eV to 1.2 keV due to the uncertainty of incident neutron flux, resolution function, background subtraction, etc., in this measurement.
The measured capture yields together with the SAMMY fits in neutron energy ranges below 100 eV are shown in Fig. \ref{Fig3}. Good agreements were observed between present data and SAMMY fits both in the resonance energy and shape.

\begin{figure}[h!]
\centering
\includegraphics[scale=0.5]{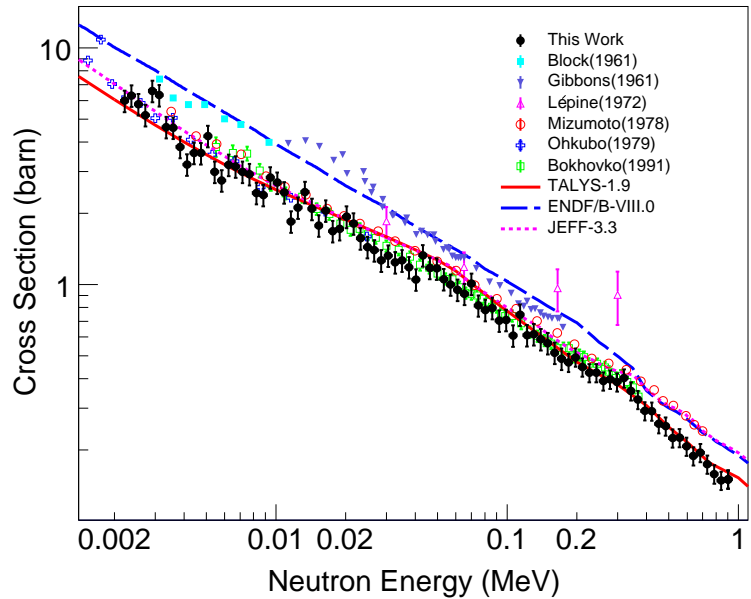}
\caption{\footnotesize (color online) Comparisons of the measured  neutron capture cross section on $^{159}$Tb of this work (black full circles) with previous measurements~\cite{Block1961,Gibbons1961,Lpine1972,Mizumoto1978,Ohkubo1979,Bokhovko1991} (symbols), evaluations~\cite{JEFF,Brown2018} (dashed lines) and TALYS-1.9 calculations~\cite{TALYS} in the 2 keV to 1 MeV energy region.}
\label{Fig4}
\end{figure}

The measured neutron capture cross section ratio of $^{159}$Tb relative to the gold standard was obtained in the neutron energy range of 2 keV to 1 MeV with the same method of Refs.~\cite{Wisshak2006,Tessler2015}. The absolute cross sections were converted from these ratios using the gold data of JEFF-3.3 evaluations. In Fig. \ref{Fig4}, the experimental $^{159}$Tb(n, $\gamma$) cross section and TALYS-1.9~\cite{TALYS} calculation result of this work are compared with the available measured data in literature~\cite{Block1961,Gibbons1961,Lpine1972,Mizumoto1978,Ohkubo1979,Bokhovko1991} and evaluated data in libraries JEFF-3.3~\cite{JEFF}, ENDF/B-VIII.0~\cite{Brown2018}. As illustrated, the present experimental results are in good agreement with the data from TALYS-1.9, Bokhovko et al.~\cite{Bokhovko1991}, and Ohkubo et al.~\cite{Ohkubo1979} in the overlapping energy regions. The cross section of JEFF-3.3 and Mizumoto et al.~\cite{Mizumoto1978} are best fitted by the present data below 400 keV neutron energy and slightly larger above 400 keV. While, the data of Block et al.~\cite{Block1961}, Gibbons et al.~\cite{Gibbons1961}, L\'{e}pine et al.~\cite{Lpine1972} and ENDF/B-VIII.0 library significantly overestimated result of this work.

The Maxwellian-averaged cross sections (MACS) for stellar temperature can be calculated from the neutron capture cross sections using the formula (\ref{eq7})~\cite{Wisshak1990,Lederer2019}:
\vspace{0.1cm}
\begin{normalsize}
\begin{equation}
MACS = \frac{2}{\sqrt{\pi}(kT)^2}\int_{0}^{\infty}{\sigma(E)Ee^{-E/kT}}dE
\label{eq7}
\end{equation}
\end{normalsize}
where the $kT$ is stellar temperature, $\sigma(E)$ is the neutron capture cross section, $E$ is the neutron energy.

\begin{table}[h!]\small
\centering
\caption{\small The contributions from JEFF-3.3 evaluated data to present MACS.}
\label{tab31}
\renewcommand\arraystretch{0.9}
\setlength{\tabcolsep}{2mm}{
\begin{tabular}{lllllllllll}
\toprule
\hline
kT (keV) &5 &10 &20 &30 &40 &50 &60 &70 &80 &100\\\hline
Ratio (\%) &10.0 &4.0 &1.5 &0.85 &0.56 &0.41 &0.32 &0.28 &0.21 &0.15\\
\bottomrule
\hline
\end{tabular}
}
\vspace{0.1cm}
\end{table}

\begin{figure}[h!]
\vspace{-10pt}
\centering
\includegraphics[scale=0.75]{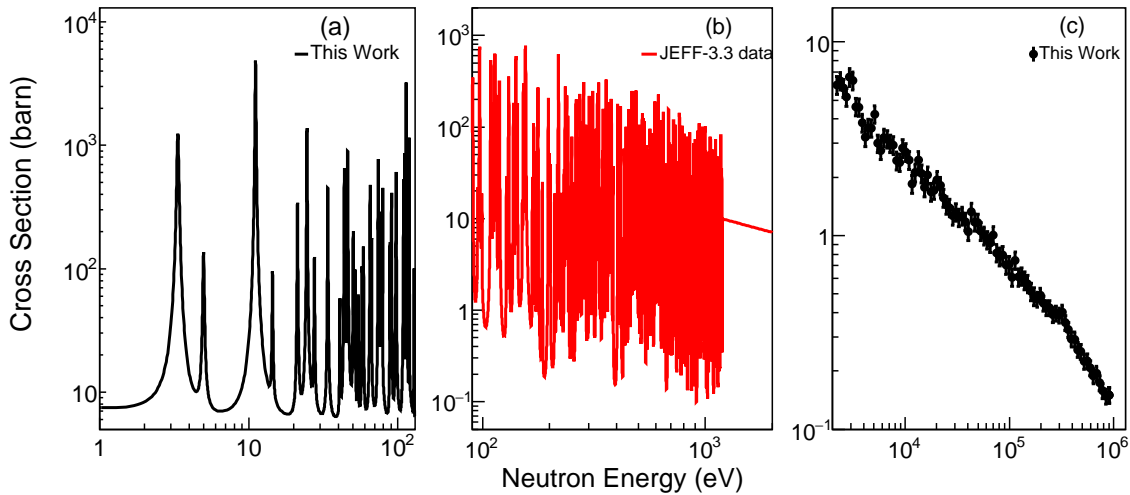}
\caption{\footnotesize (color online) The $^{159}$Tb(n, $\gamma$) cross sections used in the calculation of present MACS.}
\label{Fig5}
\end{figure}

For $^{159}$Tb(n, $\gamma$), the present MACS were determined using three parts of cross sections from 1 eV to 1 MeV energy range. The first part, shown in Fig. \ref{Fig5}(a), is the cross sections reconstructed from SAMMY fits to the present measured neutron capture yield in the resolved resonance region below 100 eV. The second part, shown in Fig. \ref{Fig5}(b), is derived from the JEFF-3.3 evaluated cross section data between 100 eV and 2 keV and its contributions to the whole MACS are found in Table \ref{tab31}. As we can see the MACS of JEFF-3.3 data cannot be ignored in the current study of s-process nucleosynthesis. The last part is provided by the present experimental cross sections relative to the gold standard over the energy range of 2 keV to 1 MeV as shown in Fig. \ref{Fig5}(c).

The result of this work at $kT$ = 30 keV are compared with the previous experiments, theoretical calculations, and recommended data from KADoNis database in Table \ref{tab3}. 
The MACS recommended data from KADoNiS-v0.3 and calculations of TALYS-1.9 code  agree with the present measured data less than $4\%$. 
The values from recent activation measurement of Praena et al.~\cite{Praena2014}, and the earlier data of Allen et al.~\cite{Allen1971}, L\'{e}pine et al.~\cite{Lpine1972} are relatively larger than the present result, whereas the result of Bokhovko et al.~\cite{Bokhovko1991} and NON-SMOKER~\cite{Rauscher2000} is smaller. Theoretical calculations from MOST 2002~\cite{Goriely2002} and MOST 2005~\cite{Goriely2005} obviously overestimated and underestimated the present experimental data by $89\%$ and $28\%$, respectively.

\begin{figure}[h!]
\centering
\includegraphics[scale=0.45]{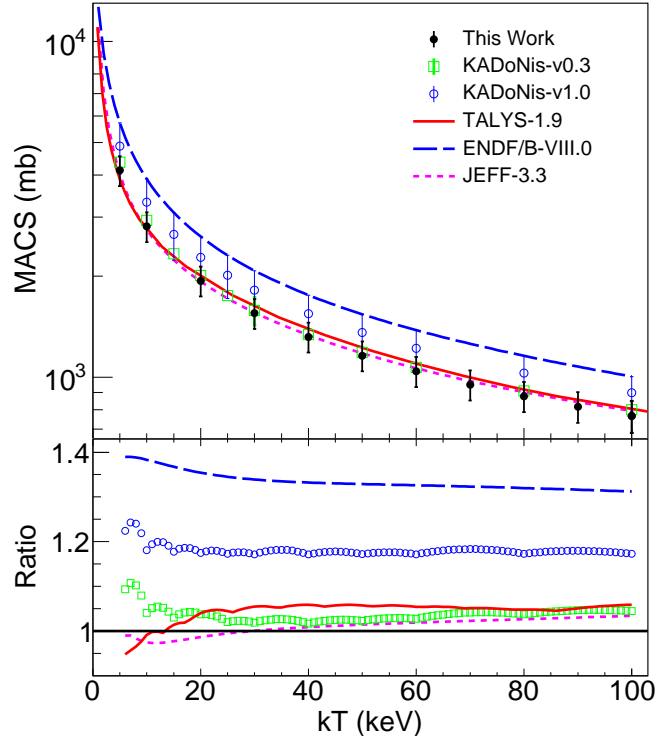}
\caption{\footnotesize (color online) (top) The MACS of $^{159}$Tb(n, $\gamma$) and (bottom) the ratios  between present experimental data and others~\cite{Dillmann,kadonisV10,Brown2018,JEFF}.}
\label{Fig6}
\end{figure}

\begin{table}[h!]\small
\centering
\caption{\small The present MACS (in mb) of $^{159}$Tb(n, $\gamma$) at kT = 30 keV compared to past experiments, calculations and evaluations.}
\label{tab3}
\renewcommand\arraystretch{1.1}
\setlength{\tabcolsep}{4.5mm}{
\begin{tabular}{llllll}
\toprule
\hline
This work  & 1560$\pm$160 & KADoNis-v0.3~\cite{Dillmann}   & 1580$\pm$150 & KADoNiS-v1.0~\cite{kadonisV10} & 1817$\pm$258 \\
Praena 2014~\cite{Praena2014}  & 2166$\pm$181 & Bokhovko 1992~\cite{Bokhovko1991} & 1471$\pm$66 &  L\'{e}pine 1972~\cite{Lpine1972}  & 1850$\pm$250 \\ 
Allen 1971~\cite{Allen1971} & 2200$\pm$200  & MOST 2005~\cite{Goriely2005} &1116 & MOST 2002~\cite{Goriely2002} &2949 \\
NON-SMOKER~\cite{Rauscher2000} & 1427 & TALYS-1.9~\cite{TALYS}    & 1628 & &\\ 
\bottomrule
\hline
\end{tabular}
}
\end{table}
\vspace{8pt}

The MACS values should be known as a function of thermal energies between a few to hundred keV at the various s-process sites.
For completeness, the comparisons of experimental MACS and TALYS-1.9 calculation result of this work to recommended data of KADoNis and evaluation libraries for stellar thermal energy from 5 to 100 keV are shown in Fig. \ref{Fig6}. On top panel: the black full circles and red solid line denote the present measured data and TALYS-1.9 calculation, respectively; the open green squares and blue circles indicate KADoNis-v0.3 and KADoNis-v1.0 recommended data; the blue dashed line and pink dotted line refer to calculated MACS by using the evaluated cross section data of ENDF/B-VIII.0 and JEFF-3.3 libraries. On bottom panel: the ratios of measured MACS to those from  KADoNis-v0.3 (green open squares), KADoNis-v1.0 (blue open circles), ENDF/B-VIII.0 (blue dashed line), TALYS-1.9 (red solid line), and JEFF-3.3 (pink dotted line) data; Black solid line corresponds to ratio 1. 
Our measured MACS data is well reproduced by the TALYS-1.9 calculation and JEFF-3.3 evaluation less than $5\%$ for all given thermal energies. While, the present MACS is remarkably smaller than those of ENDF/B-VIII.0 evaluation by 30$\%$ - 40$\%$.
Specifically, the present experimental MACS is smaller than KADoNiS-v0.3 data by about 3$\%$ - 11$\%$ and 4$\%$ for the energy region $kT < $ 20 keV and 20 keV $< kT < $ 100 keV, respectively. MACS data derived from KADoNis-v1.0 is larger about 15$\%$ - 25$\%$ than this work in the whole energy range.

The total experimental uncertainties of MACS include statistical and systematic contributions as shown in Table \ref{tab4}. The statistical uncertainty was less than 1$\%$ below 100 eV and less than 5$\%$ between 2 keV and 1 MeV. The systematic uncertainties are due to neutron flux (4.5$\%$ below 150 keV and 8.0$\%$ above 150 keV), saturated resonance peak normalization (1$\%$), background subtraction with filters (8.6$\%$), flight path (0.08$\%$, neglected), PHWT calculation (3$\%$) and sample impurity (0.01$\%$). Additionally, the contributions from neutron inelastic to present MACS are included at different stellar temperatures as shown in Table \ref{tab5}, which cannot be eliminated in the present experiment.

\begin{table}[h!]\small
\centering
\setlength{\abovecaptionskip}{0.2cm}
\setlength{\belowcaptionskip}{0.1cm}
\caption{\small Systematic uncertainties of MACS $^{159}$Tb (n, $\gamma$) cross sections.}
\label{tab4}
\renewcommand\arraystretch{0.95}
\setlength{\tabcolsep}{10mm}{
\begin{tabular}{ll}
\toprule
\hline
Source & Uncertainty(\%)\\\midrule
Neutron flux ($<$150 keV; $>$150 keV)	&4.5; 8.0\\
Pulse height weighting functions	&3.0\\
Background subtraction with filters	&8.6\\
Normalization factor             	&1.0\\
flight path                         &0.08\\
Sample impurities	                &0.01\\
Total (without inelastic contribution)    &10.2;12.2\\
\bottomrule
\hline
\end{tabular}
}
\end{table}

\begin{table}[h!]\small
\centering
\caption{\small The contributions from neutron inelastic to present MACS.}
\label{tab5}
\renewcommand\arraystretch{0.9}
\setlength{\tabcolsep}{1.5mm}{
\begin{tabular}{lllllllllll}
\toprule
\hline
kT (keV) &5 &10 &20 &30 &40 &50 &60 &70 &80 &100\\\hline
Ratio (\%) &10$^{-21}$ &10$^{-11}$ &10$^{-5}$ &10$^{-3}$ &10$^{-2}$ &0.21 &0.68 &1.61 &3.11 &7.90\\
\bottomrule
\hline
\end{tabular}
}
\end{table}

The astrophysical reaction rates at a given thermal temperature can be numerically calculated from corresponding MACS by using eq.(\ref{eq8}):
\begin{normalsize}
\begin{equation}
N_{A}<\sigma\nu>(kT) = N_{A}\times MACS \times\nu
\label{eq8}
\end{equation}
\end{normalsize}
where N$_{A}$ is the Avogadro number, MACS is defined by Equation (\ref{eq7}), $\nu =\sqrt{2kT/\mu}$ is the mean thermal velocity. The $^{159}$Tb(n, $\gamma$) rate of this work is fitted as a function of temperature T$_{9}$ (in unit of 10$^{9}$ K) with the standard form of REACLIB:
\begin{normalsize}
\begin{equation}
N_{A}<\sigma\nu>_{(n,\gamma)} = exp(16.7928 + 0.0115T^{-1}_{9} - 1.1836T^{-1/3}_{9} + 4.0069T^{1/3}_{9} -0.5127T_{9} + 0.03145T^{5/3}_{9} - 1.3577\ln T_{9})
\label{eq9}
\end{equation}
\end{normalsize}
with the fitting errors are less than 1$\%$ in the range from T$_{9}$ = 0.02 to T$_{9}$ = 10.

\section{IV. Astrophysical Implications}
In TP-AGB stars, the $^{13}$C pocket was formed following the third dredge-up (TDU) via $^{12}$C(p,$\gamma$)$^{13}$N($\beta^{+}\nu)^{13}$C and generate neutrons by reaction $^{13}$C($\alpha$,n) that trigger the s-process~\cite{Rolfs1988,Cristallo2009,Karakas2010}. The impact of $^{159}$Tb stellar neutron capture cross sections on main s-process nucleosynthesis was investigated using the stellar evolution code MESA~\cite{Paxton2011}. The stellar structure evolution and the s-process nucleosynthesis are computed with the MESA \emph{star} and \emph{net} modules separately, hence requiring less computing time and resources. A low mass star with initial mass 2$M_\odot$ and metallicity Z = 0.01 evolved from the pre-main sequence to near the end of the TP-AGB phase with nuclear network agb.net. In the $^{13}$C pocket formed after the 3rd TDU episodes, the temperature varies within 5.5 $\sim$ 8.9$\times$10$^{7}$ K and the density increase from 96.0 g$/$cm$^{3}$ to 7.7$\times$10$^{3}$ g$/$cm$^{3}$ during the whole interpulse phase.

To treat the s-process, detailed nucleosynthesis calculations were performed with MESA \emph{net} module. The network include 775 isotopes and 7344 reactions, and the temperature, density and light nuclei initial abundance were extracted from $^{13}$C pocket of the precalculated stellar structure. The reaction rates corresponding to all proton, $\alpha$, $\beta$-decay, and neutron capture are provided by JINA REACLIB data tables, based on the 2017 updated version of the compilation by Cyburt et al.~\cite{JINA2010}. Furthermore, the initial element abundances were taken from Lodders et al.~\cite{Lodders2003} for both structure and nucleosynthesis simulations.

\begin{figure}[h!]
\centering
\includegraphics[scale=0.6]{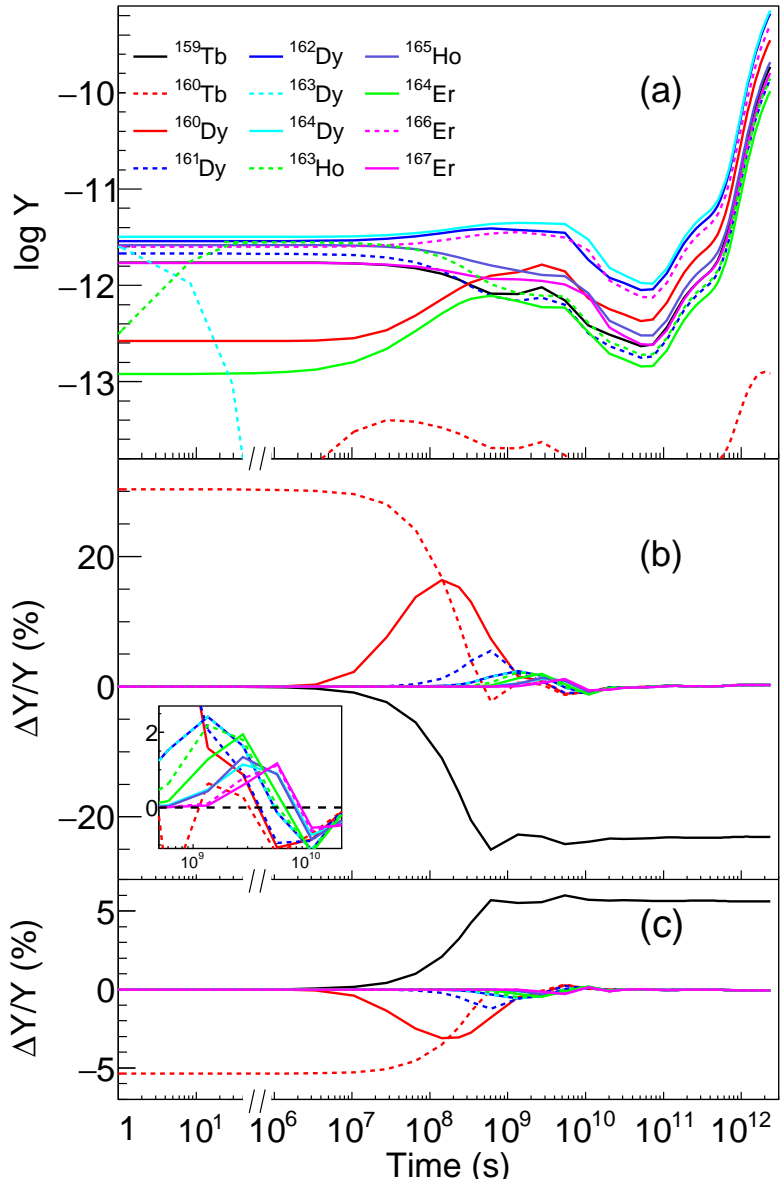}
\caption{\footnotesize (color online) (Top) Time evolution of nuclear abundances on the test trajectory with T = 8.9$\times$10$^{7}$ K and $\rho$ = 7.7$\times$10$^{3}$ g$/$$cm^{3}$. (Middle) The abundance differences between the default case of the JINA REACLIB rate for $^{159}$Tb(n, $\gamma$)$^{160}$Tb, which is derived from KADoNiS-v0.3, and the case of ENDF/B-VIII.0. (Bottom) Same physical quantities as in Middle one but the reaction rates of this work
instead of ENDF/B-VIII.0 data are used in calculations.}
\label{Fig7}
\end{figure}

We estimated the sensitivity of nuclear yields to the reaction rate of $^{159}$Tb(n, $\gamma$)$^{160}$Tb.
Fig. \ref{Fig7}(a) shows the time evolution of abundances of nuclides around A $\approx$ 160 along the s-process reaction path at constant temperature (T = 8.9$\times$10$^{7}$ K) and density ($\rho$ = 7.7$\times$10$^{3}$ g$/$cm$^{3}$) with the default case of the JINA REACLIB rate for $^{159}$Tb(n, $\gamma$)$^{160}$Tb, which is derived from KADoNiS-v0.3 database. 
The abundance of $^{163}$Dy, which has half-lives of about 50 days in stellar environment~\cite{Takahashi1987}, decrease rapidly due to considering $\beta$-decay process to $^{163}$Ho. 
The abundances of seed nuclei with initial values around A $\approx$ 160 decreased remarkably by neutron capture reactions before evolving to 7$\times$10$^{10}$ because of the s-process flow from the dominant seed nuclei $^{56}$Fe not yet overcomes the first s-process peak nuclei Sr and Ba, which act as the bottle necks in the reaction flow. While, the abundances of $^{160}$Dy and $^{164}$Er increased obviously with the depletion of $^{160}$Tb and $^{163}$Ho from about 10$^{7}$ $\sim$ 4$\times$10$^{9}$ s. Then, the productions of all nuclides around A $\approx$ 160 increased continuously up to 2 $\sim$ 3 orders larger than initial values at the end of interpulse.
Fig.\ref{Fig7}(b) shows the abundance changes between the KADoNiS-v0.3 and ENDF/B-VIII.0 rate of $^{159}$Tb(n, $\gamma$)$^{160}$Tb used in the network calculations. In this case, the data of KADoNiS-v0.3 is smaller about 40$\%$ than the one of ENDF/B-VIII.0. The abundances of $^{159,160}$Tb, $^{160,161}$Dy obviously affected by the uncertainty of cross section of $^{159}$Tb(n, $\gamma$)$^{160}$Tb reaction. While, this propagation effect less important for those nuclei further from the $^{159}$Tb. This trend is because of achieving s-process reaction flow equilibriums~\cite{Kappeler2011,Reifarth2014,Koloczek2016} due to the high neutron fluence in the $^{13}C$ pocket.
Fig.\ref{Fig7}(c) shows the same physical quantities as in Fig.\ref{Fig7}(b) but the $^{159}$Tb(n, $\gamma$)$^{160}$Tb rate of this work instead of ENDF/B-VIII.0 data are used in calculations. As given in figure, the changes of abundances are much smaller than the one of Fig.\ref{Fig7}(b) because of good agreement between two rates within 5$\%$.

\section{V. SUMMARY}
We experimentally measured $^{159}$Tb(n, $\gamma$) cross sections up to 1 MeV neutron energy by TOF method using the C$_{6}$D$_{6}$ detectors at CSNS Back-n facility. Resonance capture kernels were determined from 1 eV to 1.2 keV by analyzing the measured data with the R-matrix code SAMMY. The cross sections were obtained relative to the gold standard over the energy range of 2 keV to 1 MeV and well predicted by the TALYS-1.9 theoretical calculations and JEFF-3.3 evaluation library within experimental errors. While, the ENDF/B-VIII.0 evaluated cross section largely overestimated the present data.
Maxwellian-averaged neutron capture cross sections (MACS) extracted from the present (n, $\gamma$) data are in good agreement with the data of JEFF-3.3 and KADoNiS-v0.3 ~\cite{Dillmann} within 5$\%$ and 3$\%$-11$\%$ of discrepancies in the whole energy range, respectively. However, the data of ENDF/B-VIII.0 and KADoNiS-v1.0 ~\cite{kadonisV10} significantly overestimate the experimental ones by 40$\%$ and 20$\%$.
The sensitivity analysis of $^{159}$Tb(n, $\gamma$)$^{160}$Tb reaction rate was investigated for the stellar evolution and nucleosynthesis of the 2$M_\odot$ star model using the MESA code.
Significant changes in abundances of $^{159,160}$Tb, $^{160,161}$Dy were observed by comparing the stellar reaction rates of $^{159}$Tb(n, $\gamma$)$^{160}$Tb from KADoNiS-v0.3, this work and ENDF/B-VIII.0 evaluation. The present results show very small propagation to the more heavier elements for changing the rate of $^{159}$Tb(n, $\gamma$)$^{160}$Tb because of the establishment of a reaction flow equilibrium between the s-process main component nuclei. 

\section{acknowledgments}
This work was supported by the National Natural Science Foundation of China (Grants No. U2032146, 11865010, 12175152, 11765014 and 11609053) and Natural Science Foundation of Inner Mongolia (Grants No. 2019JQ01 and 2018MS01009). We also thank the efforts of the CSNS Back-n collaboration.
\appendix
\section{Appendix A. Resonance Parameters}
The resonance parameters obtained in this work are compared with those from JEFF-3.3 evaluations in Table A1. The experimental kernels $k$ $( k = g \Gamma_n \Gamma_\gamma/(\Gamma_n + \Gamma_\gamma)$, $g$ is the statistical factor) highly agree with the evaluations below 100 eV, while some differences were observed in the energy range of 100 eV to 1.2 keV due to the uncertainty of incident neutron flux, resolution function, background subtraction, etc., in this measurement.

\begin{center}
\tablefirsthead{%
    \multicolumn{12}{l}{\textbf{Table A.1}}\\
    \multicolumn{12}{l}{The resonance kernels in this work ($k_1$) are compared with the JEFF-3.3 ($k_2$).}\\
\hline}
\tablehead{%
	\multicolumn{12}{l}{\textbf{Table A.1}(continued)}\\
	\hline
    $E_R$(eV) &$k_1$ &$k_2$& $E_R$(eV) &$k_1$ &$k_2$  &$E_R$(eV) &$k_1$ &$k_2$  &$E_R$(eV) &$k_1$ &$k_2$\\\hline}
\tabletail{%
	\hline
	\multicolumn{12}{r}{\small\sl continued on next page}\\
    }

\tablelasttail{\hline}
\renewcommand\arraystretch{1.9}
\setlength{\tabcolsep}{0.7mm}\small{
\begin{supertabular}{lll|lll|lll|lll}
$E_R$(eV) &$k_1$ &$k_2$& $E_R$(eV) &$k_1$ &$k_2$ &$E_R$(eV) &$k_1$ &$k_2$  &$E_R$(eV) &$k_1$ &$k_2$\\\hline
3.34$\pm$0.03   & 0.16$\pm$0.01  & 0.21  & 495.58$\pm$2.18   & 10.62$\pm$2.69 & 37.82
&4.97$\pm$0.13   & 0.03$\pm$0.01  & 0.03  & 499.21$\pm$3.03   & 8.99$\pm$2.87  & 12.96\\
11.04$\pm$0.03  & 2.97$\pm$0.10  & 4.46  & 504.15$\pm$1.35   & 13.43$\pm$1.6  & 6.55
&14.40$\pm$0.30  & 0.04$\pm$0.01  & 0.07  & 511.47$\pm$1.65   & 4.77$\pm$0.695 & 14.56\\
21.19$\pm$0.13  & 0.82$\pm$0.07  & 0.71  & 518.43$\pm$3.54   & 5.23$\pm$1.87  & 12.95
&24.53$\pm$0.07  & 2.29$\pm$0.09  & 3.18  & 521.88$\pm$1.58   & 25.19$\pm$2.91 & 3.68\\
27.55$\pm$0.34  & 0.30$\pm$0.05  & 0.52  & 529.19$\pm$1.47   & 17.40$\pm$2.07 & 21.44
&33.83$\pm$0.12  & 2.15$\pm$0.12  & 1.59  & 533.60$\pm$0.88   & 14.93$\pm$2.81 & 19.16\\
40.81$\pm$0.63  & 0.21$\pm$0.06  & 0.31  & 544.84$\pm$2.70   & 4.07$\pm$1.42  & 18.45
&43.69$\pm$0.13  & 2.60$\pm$0.14  & 3.48  & 547.17$\pm$2.92   & 5.38$\pm$1.75  & 9.64\\
46.04$\pm$0.14  & 4.46$\pm$0.23  & 7.72  & 553.36$\pm$2.92   & 8.91$\pm$1.85  & 1.9
&50.16$\pm$0.24  & 1.11$\pm$0.10  & 1.17  & 558.28$\pm$4.99   & 4.80$\pm$2.44  & 8.88\\
51.61$\pm$0.38  & 0.70$\pm$0.12  & 0.52  & 561.60$\pm$3.53   & 12.98$\pm$4.92 & 7.7
&54.08$\pm$0.53  & 0.16$\pm$0.04  & 0.31  & 566.04$\pm$2.01   & 33.91$\pm$5.72 & 8.11\\
57.45$\pm$0.36  & 0.77$\pm$0.10  & 0.81  & 571.47$\pm$2.78   & 10.74$\pm$2.55 & 16.34
&58.77$\pm$0.29  & 0.85$\pm$0.10  & 0.98  & 576.44$\pm$2.79   & 2.97$\pm$1.02  & 12.92\\
65.17$\pm$0.18  & 4.57$\pm$0.25  & 6.97  & 578.42$\pm$3.35   & 4.06$\pm$1.49  & 4.59
&66.56$\pm$0.39  & 0.67$\pm$0.13  & 1.27  & 581.01$\pm$2.24   & 5.56$\pm$1.32  & 2.77\\
73.83$\pm$0.14  & 4.33$\pm$0.19  & 5.99  & 592.31$\pm$3.82   & 23.21$\pm$6.52 & 14.88
&76.45$\pm$0.29  & 2.59$\pm$0.27  & 2.44  & 594.83$\pm$2.48   & 24.69$\pm$8.43 & 17.81\\
77.98$\pm$0.37  & 1.80$\pm$0.24  & 2.53  & 597.47$\pm$3.25   & 16.48$\pm$6.58 & 35.43
&78.80$\pm$0.67  & 0.39$\pm$0.18  & 0.97  & 600.41$\pm$2.16   & 17.77$\pm$5.24 & 19.04\\
88.44$\pm$0.42  & 2.13$\pm$0.26  & 2.01  & 603.10$\pm$2.68   & 9.54$\pm$3.21  & 11.05
&90.60$\pm$0.33  & 6.59$\pm$0.67  & 3.97  & 605.90$\pm$3.19   & 4.63$\pm$1.68  & 13.72\\
97.24$\pm$0.31  & 7.58$\pm$0.93  & 10.4  & 610.50$\pm$3.57   & 16.60$\pm$3.29 & 2.31
&109.24$\pm$0.22 & 8.24$\pm$0.34  & 8.19  & 615.81$\pm$0.94   & 27.31$\pm$9.28 & 8.64\\
111.56$\pm$0.27 & 3.04$\pm$0.21  & 2.76  & 619.30$\pm$2.74   & 7.33$\pm$1.91  & 11.09
&113.89$\pm$0.26 & 6.79$\pm$0.47  & 10.66 & 627.47$\pm$3.03   & 9.57$\pm$3.67  & 22.74\\
115.73$\pm$0.32 & 2.25$\pm$0.17  & 3.62  & 630.28$\pm$3.80   & 17.25$\pm$4.37 & 8.11
&118.83$\pm$0.28 & 2.45$\pm$0.23  & 5.63  & 639.30$\pm$3.02   & 11.98$\pm$2.34 & 8.15\\
128.46$\pm$0.25 & 0.63$\pm$0.05  & 0.53  & 644.79$\pm$3.11   & 11.45$\pm$3.57 & 7.86
&137.28$\pm$0.62 & 1.28$\pm$0.27  & 1.45  & 649.11$\pm$1.49   & 23.02$\pm$5.35 & 31.82\\
138.25$\pm$0.45 & 2.96$\pm$0.36  & 3.57  & 658.22$\pm$3.85   & 1.07$\pm$0.358 & 2.46
&141.47$\pm$0.26 & 14.80$\pm$2.9  & 13.63 & 660.53$\pm$1.76   & 4.01$\pm$0.684 & 8.03\\
143.69$\pm$0.26 & 2.55$\pm$0.94  & 3.98  & 663.24$\pm$1.33   & 4.11$\pm$0.636 & 5.29
&152.90$\pm$0.24 & 9.67$\pm$0.74  & 7.38  & 678.45$\pm$2.51   & 4.75$\pm$1.06  & 4.55\\
155.95$\pm$0.28 & 6.08$\pm$0.48  & 13.48 & 681.88$\pm$2.86   & 7.41$\pm$2.1   & 16.06
&167.96$\pm$0.40 & 0.83$\pm$0.08  & 0.77  & 684.57$\pm$2.23   & 12.86$\pm$2.91 & 34.2\\
170.13$\pm$0.30 & 3.50$\pm$0.29  & 3.1   & 687.71$\pm$2.97   & 18.41$\pm$4.44 & 14.48
&173.03$\pm$0.38 & 2.07$\pm$0.14  & 1.5   & 694.55$\pm$4.29   & 8.90$\pm$2.98  & 7.61\\
177.21$\pm$0.40 & 6.40$\pm$0.45  & 8.37  & 702.00$\pm$3.12   & 15.57$\pm$3.21 & 14.2
&186.00$\pm$0.53 & 0.47$\pm$0.04  & 0.8   & 708.96$\pm$2.52   & 12.70$\pm$2.27 & 6.79\\
197.78$\pm$0.44 & 6.73$\pm$0.41  & 7.45  & 720.87$\pm$1.91   & 8.45$\pm$2.84  & 33.22
&200.24$\pm$0.48 & 1.07$\pm$0.14  & 1.81  & 723.53$\pm$0.40   & 3.99$\pm$2.51  & 1.94\\
201.84$\pm$0.02 & 0.83$\pm$0.44  & 1.39  & 727.88$\pm$4.88   & 3.51$\pm$1.75  & 6.11
&211.15$\pm$0.54 & 0.73$\pm$0.06  & 0.71  & 732.89$\pm$2.91   & 27.50$\pm$4.62 & 25.83\\
219.61$\pm$0.39 & 12.13$\pm$0.89 & 28.42 & 740.27$\pm$5.74   & 6.82$\pm$2.59  & 9.01
&226.33$\pm$0.78 & 0.97$\pm$0.11  & 0.79  & 748.01$\pm$3.48   & 4.48$\pm$1.44  & 7.29\\
229.15$\pm$1.60 & 0.41$\pm$0.10  & 0.38  & 750.91$\pm$3.58   & 4.70$\pm$1.61  & 10.19
&235.67$\pm$0.57 & 15.15$\pm$0.93 & 13.25 & 754.69$\pm$3.68   & 5.91$\pm$1.77  & 5.15\\
239.34$\pm$0.43 & 14.12$\pm$1.43 & 11.02 & 767.45$\pm$2.14   & 57.03$\pm$7.81 & 40.77
&242.17$\pm$0.45 & 5.41$\pm$0.52  & 5.98  & 768.77$\pm$6.80   & 25.42$\pm$22.3 & 19.03\\
245.28$\pm$0.16 & 0.15$\pm$0.04  & 0.47  & 783.42$\pm$4.63   & 6.08$\pm$2.81  & 3.92
&251.93$\pm$0.61 & 1.90$\pm$0.18  & 3.77  & 787.44$\pm$2.68   & 37.04$\pm$8.35 & 28.37\\
254.75$\pm$0.58 & 0.79$\pm$0.10  & 1.23  & 790.76$\pm$2.32   & 21.69$\pm$5.76 & 6.31
&263.24$\pm$0.71 & 6.89$\pm$0.52  & 10.68 & 802.50$\pm$4.95   & 12.41$\pm$3.37 & 23.45\\
269.05$\pm$0.61 & 5.16$\pm$0.33  & 4.15  & 810.17$\pm$5.36   & 10.99$\pm$5.99 & 16.41
&273.83$\pm$0.58 & 7.15$\pm$0.55  & 13.24 & 814.44$\pm$7.91   & 15.65$\pm$6.87 & 5.22\\
280.05$\pm$0.96 & 1.82$\pm$0.34  & 1.22  & 824.44$\pm$2.35   & 35.91$\pm$3.31 & 22.62
&281.86$\pm$0.69 & 6.61$\pm$0.80  & 6.71  & 837.62$\pm$4.69   & 2.01$\pm$0.79  & 3.88\\
284.77$\pm$0.53 & 17.03$\pm$3.04 & 20.96 & 845.01$\pm$3.69   & 39.72$\pm$7.69 & 25.67
&290.57$\pm$0.59 & 3.68$\pm$0.36  & 6.51  & 849.34$\pm$2.22   & 31.80$\pm$6.76 & 26.82\\
301.43$\pm$0.61 & 16.92$\pm$1.03 & 14.29 & 853.81$\pm$3.24   & 19.32$\pm$5.49 & 6.53
&306.22$\pm$0.65 & 10.40$\pm$0.81 & 9.36  & 859.26$\pm$4.33   & 8.40$\pm$3.06  & 8.88\\
312.88$\pm$0.65 & 14.52$\pm$0.97 & 17.66 & 872.33$\pm$19.42  & 3.24$\pm$2.78  & 4.12
&316.27$\pm$0.99 & 1.73$\pm$0.32  & 2.12  & 876.21$\pm$6.60   & 14.09$\pm$5.05 & 5.39\\
324.12$\pm$1.12 & 8.25$\pm$1.21  & 6.83  & 884.20$\pm$4.27   & 25.68$\pm$7.24 & 21.25
&326.58$\pm$1.26 & 10.49$\pm$2.57 & 5.41  & 896.76$\pm$9.85   & 6.21$\pm$5.75  & 9.01\\
329.02$\pm$0.71 & 22.49$\pm$5.41 & 19.12 & 901.55$\pm$3.01   & 25.35$\pm$7.00 & 24.29
&332.86$\pm$0.64 & 21.90$\pm$2.48 & 24.1  & 911.67$\pm$10.04  & 7.09$\pm$4.4   & 4.86\\
337.82$\pm$0.56 & 0.13$\pm$0.083 & 1.08  & 925.28$\pm$2.25   & 30.40$\pm$5.38 & 32.15
&340.49$\pm$0.59 & 0.18$\pm$0.090 & 0.95  & 935.80$\pm$2.14   & 138.65$\pm$11.8& 42.09\\
346.16$\pm$2.18 & 12.39$\pm$2.49 & 4.3   & 951.33$\pm$3.99   & 5.43$\pm$1.82  & 17.54
&348.50$\pm$1.58 & 10.67$\pm$3.24 & 13.32 & 955.25$\pm$6.16   & 2.95$\pm$1.56  & 6.37\\
350.83$\pm$1.34 & 4.58$\pm$1.19  & 5.62  & 966.97$\pm$2.56   & 21.96$\pm$6.41 & 10.9
&358.90$\pm$0.78 & 39.25$\pm$2.52 & 29.9  & 971.69$\pm$2.67   & 21.47$\pm$4.59 & 17.65\\
367.32$\pm$1.40 & 10.58$\pm$1.54 & 6.69  & 979.62$\pm$3.80   & 29.03$\pm$6.75 & 29.62
&370.07$\pm$1.79 & 6.75$\pm$2.20  & 3.67  & 994.59$\pm$2.90   & 42.17$\pm$6.68 & 43.79\\
372.77$\pm$0.88 & 22.02$\pm$3.26 & 16.48 & 998.76$\pm$5.29   & 24.14$\pm$1.07 & 20.79
&375.37$\pm$1.21 & 13.50$\pm$2.54 & 11.13 & 1005.2$\pm$78.0 & 1.36$\pm$0.14  & 5.74\\
379.40$\pm$0.82 & 18.89$\pm$1.62 & 20.49 & 1015.7$\pm$77.3 & 6.81$\pm$0.68  & 12.62
&385.42$\pm$0.82 & 2.83$\pm$0.34  & 3.4   & 1028.4$\pm$78.9 & 1.03$\pm$0.53  & 6.57\\
403.20$\pm$1.61 & 8.00$\pm$1.55  & 18.3  & 1036.8$\pm$78.8 & 0.28$\pm$0.28  & 17.27
&409.99$\pm$1.22 & 4.82$\pm$0.87  & 6.25  & 1040.6$\pm$79.0 & 0.04$\pm$0.01  & 23.26\\
429.98$\pm$2.88 & 0.57$\pm$0.22  & 0.64  & 1050.3$\pm$79.4 & 0.06$\pm$0.02  & 33.11
&433.16$\pm$1.11 & 6.32$\pm$0.69  & 6.29  & 1057.3$\pm$79.8 & 0.03$\pm$0.01  & 31.37\\
440.26$\pm$2.65 & 10.52$\pm$2.66 & 11.49 & 1068.1$\pm$79.9 & 4.79$\pm$0.46  & 14.64
&442.32$\pm$2.07 & 7.75$\pm$3.04  & 9.39  & 1097.8$\pm$81.8 & 0.17$\pm$0.02  & 24.61\\
444.97$\pm$2.60 & 5.81$\pm$1.58  & 3.42  & 1105.7$\pm$82.2 & 0.31$\pm$0.03  & 17.37
&451.34$\pm$4.06 & 5.19$\pm$2.31  & 1.92  & 1124.3$\pm$83.2 & 0.93$\pm$0.33  & 20.15\\
453.39$\pm$1.91 & 10.79$\pm$3.73 & 15.47 & 1132.4$\pm$83.6 & 0.77$\pm$0.17  & 21.64
&455.15$\pm$2.46 & 8.02$\pm$3.10  & 3.7   & 1142.8$\pm$85.0 & 2.98$\pm$0.29  & 7.42\\
458.69$\pm$1.06 & 9.02$\pm$1.20  & 13.7  & 1148.2$\pm$84.4 & 0.89$\pm$0.18  & 21.73
&463.83$\pm$0.02 & 0.26$\pm$0.25  & 1.13  & 1157.2$\pm$85.4 & 10.46$\pm$1.05 & 13.28\\
466.15$\pm$2.81 & 2.19$\pm$0.85  & 33.72 & 1172.9$\pm$135.7& 39.86$\pm$4.68 & 51.88
&475.37$\pm$1.47 & 22.85$\pm$2.47 & 12.53 & 1184.7$\pm$87.7 & 4.23$\pm$0.42  & 17.68\\
482.73$\pm$2.17 & 3.49$\pm$4.38  & 17.03 & 1192.3$\pm$86.9 & 1.14$\pm$0.11  & 15.7
&489.83$\pm$0.87 & 31.60$\pm$5.5  & 27.77 &  &  &\\
\end{supertabular}
}
\end{center}

\end{document}